\documentclass[11pt]{article}
 \usepackage{amsfonts}
 \usepackage{amssymb}
 \usepackage{amsmath,amsxtra,amsthm}

\usepackage[usenames]{color}

\usepackage[dvips]{graphicx}

\usepackage[all,matrix,arrow,curve]{xy}

\usepackage{mathrsfs}

\DeclareFontFamily{OT1}{pzc}{}
\DeclareFontShape{OT1}{pzc}{m}{it}{<-> s * [1.15] pzcmi7t}{}
\DeclareMathAlphabet{\mathpzc}{OT1}{pzc}{m}{it}

\DeclareSymbolFont{bbold}{U}{bbold}{m}{n}
\DeclareSymbolFontAlphabet{\mathbbold}{bbold}

\newtheorem{theorem}{Theorem}
\newtheorem{deff}{Definition}

\newtheorem{proposition}{Proposition}
\newtheorem{example}{Example}
\newtheorem{lemma}{Lemma}

\newtheorem{rem}{Remark}
\newtheorem*{result}{Main Result}

\newcommand{\npa}{\addtocounter{num}{1} \noindent
{\bf \arabic{section}.\arabic{num}}\;\;}
\renewcommand{\proof}{{\bf Proof.}~}

\newcommand{\mto}{\mapsto}
\newcommand{\bqa}{\begin{eqnarray}}
\newcommand\eqa {\end{eqnarray}}
\newcommand{\beq}{\begin{eqnarray}}
\newcommand{\beqn}{\begin{eqnarray}\nonumber}
\newcommand{\eeq}{\end{eqnarray}}
\newcommand{\be}{\begin{array}}
\newcommand{\ee}{\end{array}}

 \newcommand{\pt}{\partial}

   \newcommand\vf\varphi

 \newcommand{\Hom}{\mathrm{Hom}}

 \newcommand{\uHom}{\underline{\mathrm{Hom}}}

 \newcommand{\rk}{\mathrm{rk}}
 \newcommand{\md}{\mathrm{d}}

 \newcommand{\cS}{{\cal S}}

 \newcommand{\cC}{{\cal C}}
 
 \newcommand{\cD}{{\cal D}}
 \newcommand{\cM}{{\cal M}}
 
 \newcommand{\cO}{{\cal O}}
 
 \newcommand{\cT}{{\cal T}}
 \newcommand{\cQ}{{\cal Q}}

 \newcommand{\cF}{{\cal F}}

 \newcommand{\cm}{\mathpzc{m}}
 \newcommand{\cn}{\mathpzc{n}}

 \newcommand{\cs}{\mathpzc{s}}

 \newcommand{\cv}{\mathpzc{v}}

 \newcommand{\Z}{{\mathbb Z}}

\begin{document}
\author{G. Bonavolont\`a, A. Kotov}

  \def\sp{\mathfrak sp}
  \def\sll{\mathfrak sl}
  \def\g{{\mathfrak g}}
  \def\gl{{\mathfrak gl}}
  \def\su{{\mathfrak su}}
  \def\so{{\mathfrak so}}
  \def\sll{{\mathfrak sl}}
  \def\h{{\mathfrak h}}

  \def\P{{\mathbb P}}
  \def\H{\mathbb H}

   \def\a{\alpha}
   \def\b{\beta}
   \def\t{\theta}
   \def\la{\lambda}
   \def\e{\epsilon}
   \def\ga{\gamma}
   \def\de{\delta}
   \def\De{\Delta}
   \def\om{\omega}
   \def\Om{\Omega}

   \def\i{\imath}

 \def\gr{\g^{\scriptscriptstyle\mathrm{gr}}}
 \def\godd{\g_{\scriptscriptstyle 1}}
 \def\geven{\g_{\scriptscriptstyle 0}}
 \def\grodd{\gr_{\scriptscriptstyle 1}}
 \def\greven{\gr_{\scriptscriptstyle 0}}

  \def\sst{\scriptscriptstyle}

  \def\sot{{\;{\scriptstyle \otimes}\;}}
  \def\st{{\sst\times}}

  \def\df{{\sst\mathrm{def}}}

\def\scrE{\mathscr{E}}
  \def\scrF{\mathscr{F}}
  \def\scrJ{\mathscr{J}}
  \def\scrR{\mathscr{R}}

\title{Local BRST cohomology for AKSZ field theories: a global approach I.}

\date{}

\maketitle

\begin{abstract}

\vskip 1mm\noindent
 We study the Lagrangian antifield BRST formalism,
 formulated in terms of exterior horizontal forms on the infinite order jet space of graded fields for topological field theories associated to $Q$-bundles. In the case of a trivial $Q-$bundle with a flat fiber and arbitrary base, we prove that the BRST cohomology are isomorphic to the cohomology of the target space differential "twisted" by the de Rham cohomology of the base manifold. This generalizes the local result of G. Barnich and M. Grigoriev, computed for a flat base manifold.
\end{abstract}\vspace{5mm}

{\small 

}

\section{Introduction.}\label{general setting}

\vskip 3mm\noindent Horizontal forms constitute a bicomplex with respect to the BRST operator $\mathrm{\bf{s}}$ and the horizontal (or total) differential $\md_{h}$. We are interested in the study of the iterated $\mathrm{\bf{s}}$-cohomology $H^{\ast,\ast}(\mathrm{\bf{s}}|\md_{h})$ of the $\md_{h}$-cohomology groups of this bicomplex. Otherwise stated we are interested in the term $\mathrm{E}_{2}^{\ast|\ast}$ of its spectral sequence. Particularly relevant for the applications are the terms $H^{\ast,n}(\mathrm{\bf{s}}|\md_{h})$ of top horizontal forms ($n$ being the dimension of the base manifold) known as ``local BRST cohomology", i.e. the cohomology groups of $\mathrm{\bf{s}}$ in the space of local functionals. These groups control the deformation theory for gauge theories and encode classical observables, generalized symmetries and conservations laws (e.g. see \cite{Barnich}).

\vskip 3mm\noindent
Here we will adapt the formalism of local BRST cohomology to the specific setting of (topological) gauge field theories associated to flat $Q$-bundles ($\cite{AKSZ}, \cite{KS}$). Recall that a $Q$-bundle is a fiber bundle in the category of $Q$-manifolds.
In particular, a trivial $Q$-bundle over $T[1]X$ is a trivial bundle of graded manifolds
$$\eta: T[1]X\times \cM\to T[1]X,$$
where the cohomological vector field on the total space, $\cQ$, is $\eta$-related to the de Rham operator of the base. The space of graded sections $\underline{\Gamma}(\eta)$ is identified with the space of graded maps $\uHom(T[1]X,\cM)$ (\cite{DM99}).
In this case the BRST differential $\mathrm{\bf s}$ consists of the evolutionary vector field induced by $\cQ$ on the space of infinite jets of $\underline{\Gamma}(\eta)$ (see below and \cite{Barnich Grigoriev}).
The aforementioned BRST formalism has been studied in $\cite{Barnich Grigoriev}$ in the case of coordinate neighborhoods for both (graded)manifolds, the base $X$ and the target $\cM$. In these hypotheses the iterated cohomologies are the following
\begin{itemize}
\item [(i)] $H^{g,n}(\mathrm{\bf{s}}|\md_{h})$ is isomorphic to the cohomology $H^{g+n}(\mathrm{\bf{s}}+\md_{h})$ of the total BRST operator $\widetilde{\mathrm{\bf{s}}}=\mathrm{\bf{s}}+\md_{h}$ on horizontal forms of total degree $g+n$;
\item [(ii)] as a consequence of the hypothesis about the contractibility of the base space, the local BRST cohomology is isomorphic to the $Q$-cohomology of the target space functions.
\end{itemize}
These results are obtained by constructing (local) descent equations (in view of the contractibility assumption for the base $X$).
The local BRST cohomology in the case of an arbitrary connected base manifold $X$ and flat target space is given by Theorem $\ref{main_th}$ (see below).

\vskip 3mm\noindent Assume that the target $(\cM,\cQ_{\cM}):=(L=\oplus_{i\in \mathbb{Z}}L^{i},\cQ_L)$ is a $\mathbb{Z}$-graded $\mathbb{R}$-vector space of finite type, i.e. with $\mathrm{dim}_{\mathbb{R}}L_{i}<\infty$ for all $i$. More precisely we will assume $\cM$ to be a formal pointed $Q$-manifold.
In this case the space of graded maps $\uHom(T[1]X,L)$ is naturally identified with the module of differential forms on $X$ twisted by $L$; this identification suggests the following generalization: we replace the de Rham operator of the base with $Q_{\sst\mathrm{DR}}$, a (linear) homological vector field given by the $L$-twisted de Rham operator.
We prove that
\begin{result}[\bf Theorem \ref{main_th}]
 The iterated $\mathrm{BRST}$ complex for $\mathrm{AKSZ}$ field theories with arbitrary connected base manifold $X$ and target space $(L=\oplus_{i\in \mathbb{Z}}L_{i},\cQ_L)$
 has the following form:
 \beq \label{result}
 H^{g|n}(\mathrm{\bf{s}}|\md_h)\simeq \left( H^{\bullet}_{\mathrm{DR}} (X)\otimes H^{\bullet}_Q (L)\right)^{g+n}\,.
 \eeq
In other words, the local BRST cohomology are isomorphic to the $Q$-cohomology of the target space functions ``twisted" by the de Rham cohomology of $X$.
\end{result}

\noindent
An immediate interpretation for this result is the following: the $\mathrm{\bf{s}}$-cohomology in the space of local functionals contains a very restrictive information. More general functionals are needed in order to incorporate TFT (and especially those of AKSZ-type) in the frame of variational calculus for Classical Field Theory (\cite{Boc97},\cite{Beilinson-Drinfeld}). We shall continue investigating this subject in \cite{BK} by the use of different tools as ``multivalued Lagrangians" and the theory of coverings for non linear PDE-s.

\vskip 3mm\noindent
The paper has the following content. In Section $\ref{jet}$ we recall basic notions about jet spaces as the Cartan distribution, evolutionary vector fields, $\mathcal{D}$-modules, variational bicomplex and the horizontal complex. Particularly relevant for the rest of the paper will be the choice of a specific subcomplex of the horizontal complex denoted with $\bar\Omega^{\bullet}_{\sst\mathrm{poly}}(\pi)$ (see Proposition $(\ref{horizontal_polynomial})$ below).

\vskip 3mm\noindent
In Section $\ref{main}$ we construct the proof of Theorem $\ref{main_th}$ in two steps: first we prove that the local BRST cohomology ${H}^{g,n}(\mathrm{\bf{s}}|\md_h)$ are still isomorphic to the total cohomology ${H}^{g+n}(\mathrm{\bf{s}}+\md_{h})$ and then we calculate the latter cohomology by the use of an argument based on the formal integrability for a compatibility complex (see \cite{Quill},\cite{Spe69}, \cite{KV98}).
\vskip 3mm\noindent
Here we introduce some of the notations employed in the paper.
 If $\verb"M"$ is a sheaf on a manifold $X$ then $\verb"M"(U)$ is the space of its sections over an open set $U\subset X$; in the case of canonical sheaves, $X$ will appear as a subscript e.g.:  $\Omega_X$ is the sheaf of differential forms, $\cT_X$ - the sheaf of vector fields, $\cD_X$ - the sheaf of differential operators. With $\Omega (X)$  we mean the space of sections over $X$, that is, all forms; $\cT (X)$ - all vector fields, $\cD (X)$ - all differential operators.
 Analogously for the bundle of forms we write $\Lambda_X$.

\section{Jet bundles, $\cD-$modules, and local functionals.}\label{jet}

\vskip 3mm\npa In this section we review basic facts about jet spaces. Let $\pi \colon E\to X$ be a vector bundle over an $n-$dimensional smooth manifold.
Let $\mathrm{J}^k (\pi)$ be the space of $k-$jets of its sections:
 \beq
  \mathrm{J}^k (\pi)=\{ [s]_x^k \mid x\in X, \, s\in \Gamma (\pi)\}\,.
 \eeq

\vskip 1mm\noindent It is obvious that $\pi_{\sst k}\colon \mathrm{J}^k (\pi)\to X$ inherits a vector bundle structure for all $k\ge 0$, where $\pi_{\sst k}([s]_x^k)=x$.
Furthermore, there exists a canonical surjective vector bundle morphism $\pi_{\sst k,l}\colon \mathrm{J}^k (\pi)\to \mathrm{J}^l (\pi)$ for all $k\ge l$,
so that $\pi_{\sst k,l}([s]_x^k)=[s]_x^l$.
 The collection of vector bundles $\pi_{\sst k}$ together with projections $\pi_{\sst k,l}$ constitutes an inverse system, which allows to define the projective limit
 $\pi_{\sst\infty}\colon\mathrm{J}^{\infty} (\pi)\to X$, called the \textit{infinite jet space}, along with projections $\pi_{\sst\infty, k}\colon \mathrm{J}^{\infty} (\pi) \to \mathrm{J}^k (\pi)$, $k\ge 0$. The algebra of smooth functions on $\mathrm{J}^\infty (\pi)$, $\cF (\pi)$, is defined to be the direct limit of
$\cF_{k} (\pi)= C^\infty (\mathrm{J}^k (\pi))$,
\beq
\cF (\pi)\colon =\bigcup\limits_k \cF_k (\pi)\,.
\eeq

\vskip 1mm\noindent Each element of $\cF_k (\pi)$ is regarded as a \textit{nonlinear
scalar differential operator} of order $k$ acting on sections of $\pi$; this correspondence is established by the following formula:
\beq\label{function_as_nonlinear_diff}
 \cs \mto f [\cs]=j_{\sst k}(\cs)^* (f)\in C^\infty (X)\,, \hspace{3mm} \cs\in \Gamma (\pi)\, , f\in \cF_k (\pi)\,,
\eeq
where $j_{\sst k}(\cs )$ is the $k-$jet of $\cs$, regarded as a section of $\pi_{\sst k}$, so that $j_{\sst k}(\cs )(x)=[s]_x^k$.

\vskip 3mm\npa Let $\pi'\colon E'\to X$ be another bundle over the same manifold. We denote by $\cF_k (\pi, \pi')$ and $\cF (\pi, \pi')$
the space of smooth sections of the pull-back bundles
$\pi_{\sst k}^* (\pi')$ and $\pi_{\sst\infty}^* (\pi')$, respectively. Similarly to scalar functions on the space of jets, $\cF_k (\pi, \pi')$ is canonically identified with nonlinear PDEs
of maximal order $k$ acting from $\Gamma (\pi)$ to $\Gamma (\pi')$.

\vskip 3mm\npa The tangent space to $j_{\sst k-1}(\cs )$ at $x_{\sst k-1}=[\cs]_x^{k-1}$ is uniquely determined by $x_{\sst k}=[\cs]_x^{k}$; this allows to
define a vector bundle $\tau_{\sst k}\colon L^{k}\to\mathrm{J}^{k} (\pi)$, the fiber $L_{x_{\sst k}}$ of which is the tangent space to $j_{\sst k-1}(\cs )$ at $x_{\sst k-1}$.

 \vskip 3mm\begin{proposition}\label{properties_tau_bundle}  It is easy to verify that following properties hold.   
 \begin{enumerate}
  \item $\md \pi_{\sst k-1,k-2} (L_{x_k})=L_{x_{k-1}}$ for all $k\ge 2$ and $\md \pi_{\sst k-1} (L_{x_k})=T_{x} X$ for all $k\ge 1$
  \footnote{Hereafter one has
  $x_{\sst l}=\pi_{\sst k,l} (x_{k})$ for all $k\ge l$ and $x=\pi_{\sst k} (x_{k})$ for all $k\ge 0$,
  unless the contrary is expressed.}.
  \item Therefore
  $\tau_{\sst k} \simeq \pi_{\sst k}^* \left( \tau\right)$, where $\tau\colon TX\to X$ is the tangent bundle.
  \item Sections of $\tau_{\sst k}$ can be viewed as derivations of $\cF_{k-1} (\pi)$ with values in $\cF_k (\pi)$
   and sections of $\tau_{\sst \infty}=\pi_{\sst \infty}^* \left( \tau\right)$ - as derivations of $\cF (\pi)$ with values in $\cF (\pi)$, respectively.
  \item There exists a canonical bracket on $\Gamma (\tau_{\sst k})$ with values in $\Gamma (\tau_{\sst k+1})$, which gives rise to a Lie bracket
  on $\Gamma (\tau_{\sst \infty})$. The latter coincides with the commutator of the corresponding derivations of $\cF (\pi)$, hence
  $\tau_{\sst \infty}$ determines an involutive distribution $\cC(\pi)$ on $\mathrm{J}_m^\infty (\pi)$,  called the \textit{infinite Cartan distribution}.
  \item Sections of $\pi_{\sst \infty}$, which are integral leaves of $\cC (\pi)$, are of the form $j_{\sst \infty}(\cs )$ for some $\cs\in\Gamma (\pi)$.
 \end{enumerate}
 \end{proposition}

 \vskip 3mm\noindent Taking into account the above isomorphism
$\cC (\pi)\simeq\pi_{\sst\infty}^* (\tau)$, we can canonically lift any vector field on $X$ to a vector field on $\mathrm{J}^\infty (\pi)$, tangent to the Cartan distribution. Moreover, this lifting respects the Lie bracket, thus it can be viewed as a (non-linear) flat connection in $\pi_{\sst\infty}$.
The canonical lift of a vector field $\cv$ is called the \textit{total derivative} along $\cv$.
 More concretely, let $U\subset X$ be a coordinate chart together with local coordinates $\{x^i\}$ and let $\{u^a\}$,  $a=1, \ldots , \rk (\pi)$, be the linear fiber
coordinates corresponding to some trivialization of $\pi_{\sst U}$, the restriction of $\pi$ to $U$.
Let $\cv =\sum\limits_{i=1}^n h^i (x)\pt_{x^i}$ be a vector field in $U$. Then for any $f\in\cF (\pi)$,

\beq\label{total_in_coordinates}
  \bar\cv (f) =\sum\limits_{i=1}^n h^i (x) D_{x^i} f\,, \hspace{1mm} \mathrm{where}\hspace{2mm}
  D_{x^i} =
  \pt_{x^i}+\sum\limits_{a=1}^{\rk (\pi)}\sum\limits_{(\sigma)} u^a_{(\sigma\!+1_i)}\pt_{u^a_{(\sigma)}}\,.
\eeq
Here $(\sigma) =(\sigma_1, \ldots , \sigma_n)$ is a multi-index,  
$(\sigma\!+1_i)=(\sigma_1, \ldots , \sigma_i+1, \ldots, \sigma_n)$, and $\{u^a_{(\sigma)}\}$ are the fiber linear coordinates on the trivialization of $\mathrm{J}^\infty (\pi_{\sst U})$, such that the infinite jet of a section $u^a =u^a (x)$, $a=1, \ldots, \rk (\pi)$ is represented by the formula $u^a_{(\sigma)} (x)= \pt_{(\sigma)} u^a (x)$.
 Henceforth we shall use the notation $\pt_{(\sigma)}$ for
$\left(\pt_{x^1}\right)^{\sigma_1}
\ldots \left(\pt_{x^n}\right)^{\sigma_n}$ and $D_{(\sigma)}$ for
$\left(D_{x^1}\right)^{\sigma_1}
\ldots \left(D_{x^n}\right)^{\sigma_n}$, respectively.

\vskip 3mm
\begin{proposition}\label{total_property} Given any $\cv\in\cT (X)$, $\cs\in \Gamma (\pi)$, and $f\in\cF (\pi)$, one has
\beq\label{total1}
\bar\cv (f)[\cs]=\cv \left( f[\cs]\right)\,.
\eeq
\end{proposition}

\vskip 3mm\npa The Cartan distribution on $\mathrm{J}^\infty (\pi)$ allows to define an $\cF-$module of \textit{horizontal} (tangent to the Cartan distribution) vector fields
as well as an $\cF-$module of \textit{$\cC-$differential operators} $\cC\cD (\pi)$, generated by $\cC\cT (\pi)$.
 Apparently, $\cC\cT (\pi)$, as an $\cF-$module, admits a canonical (involutive) complement, consisting of $\pi_{\sst\infty}-$vertical vector fields.

\vskip 3mm
\begin{rem} The Lie subalgebra of horizontal vector fields lifted from $X$ generates $\cC\cT (\pi)$ over $\cF$.
  Along with vector fields on $X$, we can canonically lift differential operators on $X$, $\cD (X)$, to $\cC-$differential operators.
   Furthermore,  $\cC\cD (\pi)=\cF\otimes_{C^\infty (X)}\cD (X)$.
 \end{rem}

\noindent Let us consider the Lie subalgebra vector fields on $J^\infty (\pi)$, which preserve $\cC$, denoted by $\cT_{\sst\cC}(\pi)$.
Apparently, $\cC\cT (\pi)$ is an ideal in $\cT_{\sst\cC}(\pi)$. Let us define
 \beq\label{evolutionary1}
 \cT_{\sst\sf\! sym} (\pi)=\cT_{\sst\cC}(\pi) / \cC\cT (\pi)\,.
\eeq
Elements of $\cT_{\sst\sf\! sym} (\pi)$ are uniquely represented by $\pi_{\sst\infty}-$vertical vector fields which preserve $\cC$, called \textit{evolutionary vector fields}; they can be identified with sections of $\varkappa (\pi)=\pi_{\sst\infty , 0}^* (\pi)$ as follows:
 \beqn \cT_{\sst\sf\! sym} (\pi)\ni \cv\mto \phi_\cv =\cv_{\mid \cF_0}\,. \eeq

\vskip 3mm
\begin{rem}\label{evolutionary_as_derivations_over_D}
Taking into account that every total derivative is a $\pi_{\sst\infty}-$projectable vector field on $J^{\infty} (\pi)$, and thus it preserves the subspace of $\pi_{\sst\infty}-$vertical vector fields,
we immediately conclude that evolutionary vector fields are those and only those which commute with all total derivatives. In other words,
an evolutionary vector field is a derivation of $\cF (\pi)$ over $\cD (X)$. All sections of $\pi_{\sst\infty}$ which are integral leaves of the Cartan distribution, are in one-to-one correspondence with infinite jets of sections of $\pi$; therefore any infinitesimal bundle morphism of $\pi_{\sst\infty}$ preserving $\cC(\pi)$, determines an infinitesimal flow on $\Gamma (\pi)$. Hence an evolutionary vector field is a ``good candidate" for being a vector field on the space of sections. Indeed, evolutionary vector fields induce derivations of local functionals (see the later remark \ref{evolutionary_as_derivations_of_functionals}). However, almost all evolutionary vector fields, except those which come from infinitesimal morphisms of $\pi$, will not generate a flow. What concerns bundle morphisms of $\pi$, they obviously act on $\Gamma (\pi)$, so that the corresponding infinitesimal generators, which are $\pi-$projectible vector fields on the total space of $\pi$, can be thought of as ``honest" vector fields on $\Gamma (\pi)$. In other words, any $\pi-$projectible vector field $\cv$ admits the unique lift $\tilde\cv$, which preserves the Cartan distribution, that is, $\tilde\cv\in\cT_{\sst\cC}(\pi)$.
In coordinates as in (\ref{total_in_coordinates}), if
\beqn
 \cv = \sum\limits_{i=1}^n h^i (x)\pt_{x^i} +\sum\limits_{a=1}^{\rk (\pi)} g^a (x, u)\pt_{u^a}\,,
\eeq
then
\beq\label{lift_of_projectible_field}
 \tilde\cv = \sum\limits_{i=1}^n h^i (x)D_{x^i} + \sum\limits_{a=1}^{\rk (\pi)}\sum\limits_{(\sigma)} D_{(\sigma)}\left(-\sum\limits_{i=1}^n h^i u^a_i + g^a (x, u)\right)\pt_{u^a_{(\sigma)}}\,.
\eeq
One can easily check that, in contrast to total derivatives, $\tilde\cv$ preserves $\cF_k (\pi)$ for all $k$. The $\pi_{\sst\infty}-$vertical part of (\ref{lift_of_projectible_field}) is the evolutionary vector field corresponding to $\cv$.
 \end{rem}

\npa
Define the module $\Omega^i (\pi)$ of differential $i$-forms on $\mathrm{J}^{\infty} (\pi)$ by setting\footnote{Direct limit of differential forms and embeddings induced by the projections $\pi$ and $\pi_{k+1,k}$.} $$\Omega^i (\pi)\colon = \bigcup\limits_k \Omega^i (\pi_k),$$
where $\Omega^i (\pi_k)$ is the module of $i$-forms on $\mathrm{J}^{k} (\pi)$. Let us set $\Omega^* (\pi)=\oplus_{i=0}^{\infty}\Omega^i (\pi)$.\\
The decomposition of vector fields on the infinite jets space into the sum of horizontal and vertical parts
gives rise to a bicomplex structure on $\Omega^* (\pi)$,
called the \textit{variational bicomplex}:

 \beq\label{variational-bicomplex}
 \Omega^{\bullet}=\bigoplus\limits_{p,q\ge 0} \Omega^{p,q} (\pi)\,, \hspace{3mm} \md=\md_{h}+\md_{v}\,,
 \eeq
 where
  \beqn
    \xymatrix{ & \Omega^{p,q} (\pi) \ar[dl]_{\md_{h}}\ar[dr]^{\md_{v}}& \\
    \Omega^{p+1,q} (\pi) && \Omega^{p,q+1} (\pi)
               }
  \eeq
such that $ \Omega^{0,1} (\pi)$ is the annihilator of the Cartan distribution and $ \Omega^{1,0} (\pi)$ is the space of $\pi_{\sst\infty}-$horizontal
$1-$forms. In local coordinates as in (\ref{total_in_coordinates}), one has
\beq\label{variational_bicomplex_in_coordinates}
\md_{h} = \sum\limits_{i=1}^n \md x^i D_{x^i}\, , \hspace{3mm}
\md_{v} = \sum\limits_{a=1}^{\rk (\pi)}\sum\limits_{(\sigma)} \vartheta^a_{(\sigma)}\pt_{u^a_{(\sigma)}}
\eeq
where $\vartheta^a_{(\sigma)}$ are the (local) Cartan $1-$forms defined as follows:
\beq\label{Cartan_local_forms}
\vartheta^a_{(\sigma)} = \md u^a_{(\sigma)} - \sum\limits_{i=1}^n u^a_{(\sigma\!+1_i)}\md x^i\,.
\eeq

\noindent Hereafter we use the notation $\bar\Lambda^p (\pi)$ for the bundle $\Lambda^{p,0} (\pi)$ of horizontal $p-$forms
and $\left(\bar\Omega^{\bullet} (\pi), \md_h\right)$ for the
horizontal part of the variational bicomplex (\ref{variational-bicomplex}), $\left(\Omega^{{\bullet},0} (\pi), \md_h\right)$, respectively. Similarly to scalar functions, any $p-$form $\omega\in\bar\Omega^p (\pi)$ can be regarded as a nonlinear
differential operator with values in $p-$forms on $X$, acting on sections of $\pi$ by the following formula:
\beq\label{pullback_horizontal_forms}
 \cs \mto \omega [\cs]=j_{\sst k}(\cs)^* (\omega)\in \Omega^p (X)\,, \hspace{3mm} \cs\in \Gamma (\pi)\,.
\eeq
The next property is immediate from (\ref{total1}) and (\ref{variational_bicomplex_in_coordinates}):
\beq\label{total2}
(\md_h \omega) [\cs]=\md \left( \omega [\cs]\right)\,.
\eeq
By (\ref{pullback_horizontal_forms}) we conclude that, if $X$ is oriented, then any horizontal top-form $\omega\in\bar\Omega^n (\pi)$ determines a local
(that is, a jet depending) functional on
$\Gamma (\pi)$,
\beq\label{local_functional}
 \cs \mto \int\limits_X \omega [\cs]\,,
\eeq
so that, if $X$ is a compact oriented manifold without boundary then the above functional is determined by the cohomology class of $\omega$
in $H^n (\bar\Omega^{\bullet} (\pi), \md_h)$. We denote the space of local functionals by $Loc (\pi)$ and summarize the above considerations as follows.

\vskip 3mm
\begin{proposition}\label{local_functionals_as_cohomology} Let $X$ be a compact oriented manifold without boundary, then
\beqn Loc (\pi)\simeq H^n (\bar\Omega^{\bullet} (\pi), \md_h)\,.\eeq
\end{proposition}

\begin{rem}\label{evolutionary_as_derivations_of_functionals}
From the remark \ref{evolutionary_as_derivations_over_D} we conclude that any evolutionary vector field preserves the bicomplex structure (\ref{variational-bicomplex}).
 In particular, this implies that, if $X$ is compact without boundary, then, by proposition \ref{local_functionals_as_cohomology}, evolutionary vector fields are acting in $Loc(\pi)$.
 \end{rem}

\npa
For a generic fiber bundle $(E,\pi,X)$ we recall some standard results about horizontal cohomologies, see \cite{FT},\cite{Boc97},\cite{GMS}. Note that all the aforementioned results about jet spaces (e.g. Cartan distribution, variational bicomplex, etc.) can be generalized to the case of an arbitrary smooth fiber bundle.
 The exterior algebra $\Omega^\bullet(\pi)$ provides the (infinite order) de Rham complex
$$\xymatrix{0\ar[r]&\mathbb{R}\ar[r] & \Omega^{0}(\pi)\ar[r]^-{\md}&\Omega^{1}(\pi)\ar[r]^-{\md}&\ldots\\}.$$
 First we remind the following\footnote{It is based on the fact that jet bundles $\mathrm{J}^{k+1}(\pi)\to \mathrm{J}^{k}(\pi)$ are affine.}
\begin{proposition}The cohomology $H^{\ast}(\Omega^{\bullet}(\pi))$ of the previous de Rham complex is equal to the de Rham cohomology $H^{\ast}(E)$ of the total space $E$.
\end{proposition}
\noindent Recall that there is a canonical homomorphism between the de Rham cohomologies of the base and the total space
$$\pi^{\ast}:H^\ast(X)\to H^{\ast}(E);$$
if $\cs\in \Gamma(\pi)$ is a global section we denote with $\cs^{\ast}$ the corresponding epimorphism $\cs^{\ast}:H^\ast(E)\to H^{\ast}(X)$.
Whenever this epimorphism is defined, $\pi^{\ast}$ becomes a monomorphism. In this hypothesis we extend the monomorphism from the de Rham cohomology groups of the base $X$ to those for the infinite jets space
$$\pi^{\ast}:H^\ast(X)\hookrightarrow H^{\ast}(\Omega^{\bullet}(\pi)).$$
In the previous paragraph we have already introduced the splitting of $\Omega^{\bullet}(\pi)$ into horizontal and vertical parts; we denote with
$$\pi^{\bullet,0}:\Omega^{\bullet}(\pi)\to \bar\Omega^{\bullet}(\pi):=\Omega^{\bullet,0}(\pi) $$
the horizontal projection. It is obvious that this projection is a chain map
$$\md\circ \pi^{\bullet,0}=\pi^{\bullet,0}\circ \md_{h}$$
and it defines a homomorphisms of groups
$$({\pi^{\bullet,0}})^\ast: H^{\ast}(\Omega^{\bullet}(\pi))\to H^{\ast}(\bar\Omega^\bullet(\pi)).$$
The composition of the previous two cohomology maps
\beq\label{monomorphism}({\pi^{\bullet,0}})^\ast\circ \pi^\ast: H^\ast(X)\to H^{\ast}(\bar\Omega^\bullet(\pi)),\eeq
in the case $(E,\pi,X)$ admits a global section, is still a monomorphism.
It is again a well-known result ({\em loc.cit.}) the fact that $H^{\ast}(\bar{\Omega}^\bullet(\pi))$ for $\ast<n$ is equal to the de Rham cohomology of the total space $H^{\ast}(E)$.

\vskip 3mm\noindent
We shall adapt these results to our specific setting, i.e. $(E,\pi,X)$ is in particular a vector bundle.
In this case the canonical choice for the aforementioned global section is the zero section and the cohomology $H^{\ast}(E)$ coincides with the de Rham cohomology of the base $H^{\ast}(X)$.
Apparently, in our hypothesis, the horizontal cohomologies (of degree less than $n$) are provided by the image of the de Rham complex of the base, lifted by the pullback of the projection map.
In the next paragraph we will restrict our attention to the subcomplex of horizontal forms which vanish on the infinite jet of the zero section; this subcomplex is complementary to the image of the forms from the base.


\vskip 3mm \npa Among all functions on the space of $k-$jets of a (possibly graded super) vector bundle, there are two distinguished $\Z-$graded subalgebras: of fiber-wise polynomial functions, $\cS_k^{\bullet} (\pi)$,
and fiber-wise polynomial functions, vanishing on the zero section of $\pi$, $\cS_k^+ (\pi)$, which can be identified with sections of\footnote{symmetric powers of the dual bundle. } $\mathrm{Sym}^{\bullet} (\pi^*_{\sst k})$
 and $\mathrm{Sym}^+ (\pi^*_{\sst k}) =\oplus_{j>0}\mathrm{Sym}^j (\pi^*_{\sst k})$, respectively. In the case of a graded super vector bundle, the symmetric powers should be understood in the super sense.
 Given that $\pi^*_{\sst\infty}$ is a direct limit of $\pi^*_{\sst k}$,
and thus $\mathrm{Sym}^p (\pi^*_{\sst\infty})$ is a direct limit of $\mathrm{Sym}^p (\pi^*_{\sst k})$, a section of $\mathrm{Sym}^p (\pi^*_{\sst\infty})$
is always a section of $\mathrm{Sym}^p (\pi^*_{\sst k})$ for some $k$. We denote by $\cS^{\bullet} (\pi)$ and $\cS^+ (\pi)$ the direct limit of the corresponding algebras.

\begin{rem}\label{polynomials_as_multioperators} According to (\ref{function_as_nonlinear_diff}), an element of $\cS_k^p (\pi)$ can be viewed
as a symmetric $p$-linear differential
operator of maximal order $k$ acting from sections of $\pi$ to smooth functions on $X$: in order to verify this statement, we use the usual correspondence between
polynomial and symmetric multi-linear maps.
\end{rem}

\noindent From (\ref{total_in_coordinates}) one can see that the subspaces $\cS^p (\pi)$ are preserved by total derivatives for all $p$, thus we obtain
an action of $\cD (X)$ on $\cS^p (\pi)$, and finally on $\cS^{\bullet} (\pi)$ and $\cS^+ (\pi)$. In order to determine the precise form of this action, we shall first give a very brief survey of the properties of modules over $\cD_X$, the sheaf of differential operators on $X$, called \textit{$\cD-$modules}; nowadays it is a convenient language for talking about linear PDEs and their solutions. The structure sheaf of smooth functions on $X$ will be denoted with $\cO_X$  (its sections over $U$ is just $C^\infty (U)$); the choice for this convention is so motivated: many properties stated hereafter can be generalized to the analytic and algebraic case. 

\vskip 3mm\npa Denote by $\mathsf{Mod}(X)$ and $\mathsf{Mod}(X)^{\mathsf{r}}$
- the categories
 of \textit{left and right $\cD-$modules}, respectively. Eg. the structure sheaf $\cO_X$ is a left $\cD-$module, while $\Omega^n_X$, the sheaf of top degree forms on
 $X$, is a right $\cD-$module, where the right action on $\Omega^n_X$ is generated by
 \beqn
  \omega \cv = -L_\cv (\omega ) \hspace{3mm} \forall \, \omega\in\Omega^n_X\, , \cv\in\cT_X\,.
 \eeq
 Here $\cT_X$ is the sheaf of vector fields on $X$. Recall that:
  \begin{itemize}
   \item if $\verb"M"$ and $\verb"N"$ belong to $\mathsf{Mod}(X)$, then so do $\Hom (\verb"M",\verb"N")$, $\verb"M"\otimes\verb"N"$, and $\mathrm{Sym}^p(\verb"M")$ for all $p$; the symmetrization is to respect the sign rule in the super case\footnote{The bifunctors $\Hom$ and $\otimes$ are defined over $\cO_X$.}.
   \item if $\verb"M"\in\mathsf{Mod}(X)$ and $\verb"N"\in\mathsf{Mod}(X)^{\mathsf{r}}$, then $\verb"N"\otimes \verb"M"\in\mathsf{Mod}(X)^{\mathsf{r}}$,
 where
  \beqn
   (\cn \otimes \cm)\cv =  \cn \cv\otimes \cm - \cn \otimes \cv\cm\, , \hspace{3mm} \forall \, \cm\in \verb"M" \, ,\cn\in\verb"N"\, , \cv\in\cT_X\,.
  \eeq
 \item if $\verb"N"_1, \verb"N"_2\in\mathsf{Mod}(X)^{\mathsf{r}}$ then $\Hom (\verb"N"_1, \verb"N"_2)\in\mathsf{Mod}(X)$, where
  \beqn
   \cv\psi (\cn) =  \psi (\cn \cv) - \psi (\cn )\cv\, , \hspace{3mm} \forall \, \cn\in\verb"N"_1\, , \psi\in \Hom (\verb"N"_1, \verb"N"_2)\, , \cv\in\cT_X\,.
  \eeq
  \end{itemize}
 The tensor product $\otimes$ determines a symmetric monoidal structure in $\mathsf{Mod}(X)$ with $\cO_X$ as unit.
 \begin{deff} A commutative $\cD-$algebra is an algebra in the symmetric monoidal category $\left(\mathsf{Mod}(X), \otimes, \cO_X\right)$,
 i.e. a commutative monoid in the category of $\cD-$modules.
 \end{deff}
\noindent More explicitly, a commutative $\cD-$algebra is a $\cD-$module $\mathcal{A}$ together with two $\cD_X$-linear maps, (product) $$\mu:\mathcal{A}\otimes\mathcal{A}\to \mathcal{A}$$ and (unit) $$i:\cO_X\to \mathcal{A},$$ which respect the usual associativity, unitality and commutativity constraints. Note that the action of a vector field on $M$ on a product $\mu(a\otimes a')$ verifies the Leibniz' rule for any $a,a'\in \mathcal{A}$.

\vskip 3mm
\begin{example} Given any vector bundle $\pi$, $\cF (\pi)$ is a $\cD-$algebra, where the $\cD-$module structure is defined by total derivatives. Another example is the algebra of functions on an infinitely prolonged system of nonlinear partial differential equations, regarded as a ``submanifold" in $\mathrm{J}^\infty (\pi)$.
\end{example}
\begin{deff} An evolutionary vector field for a $\cD-$algebra $\mathcal{A}$ is a derivation of $\mathcal{A}$ commuting
with the action $\cD_X$.
\end{deff}

\npa Denote by $\cD (\a , \b)$ the space of linear differential operators acting between sections of vector bundles $\a$ and $\b$ on $X$, and by $\mathbbold{1}^k$
the trivial vector bundle of rank $k$. Then $\cD (\a, \mathbbold{1})$ is left $\cD-$module, which is isomorphic to $\cS^1(\pi)=\Gamma (\pi_{\sst\infty}^*)$ (see the remark \ref{polynomials_as_multioperators}); here $\cD (X)$ is acting from the left by composition. Likewise, $\cD (\a, \Lambda^n_X)$ is right $\cD-$module, where $\Lambda^p_X$
is the bundle differential $p-$forms on $X$.

\vskip 3mm
\begin{deff} Let $\a$ be a vector bundle. Denote the conjugated vector bundle $\Hom (\a, \Lambda^n_X)$ by $\hat\a$.
\end{deff}
\begin{proposition} There exists a canonical isomorphism of $\cO_X-$bimodules
$\cD (\a, \b)\simeq \cD (\hat\b, \hat\a)$, determined by formal conjugation.
In particular, $\cD (\a, \Lambda^n_X)\simeq \cD (\mathbbold{1}, \hat\a)$.
  The latter is also an isomorphism of right $\cD-$modules.
\end{proposition}

\noindent Consider the following complex of right $\cO_X-$modules $\left(\mathrm{Sym}^l\cD (\pi, \Lambda^{\bullet}_X), \md_{\sst\mathrm{DR}}\right)$,
where $\mathrm{Sym}^l\cD (\a, \b)$ is, by definition, the space the $q-$linear symmetric differential operators acting from sections of a vector bundle $\a$ to sections of another vector bundle $\b$, and the differential $\md_{\sst\mathrm{DR}}$ is induced by the left composition with the de Rham operator. The statement from Remark \ref{polynomials_as_multioperators} about polynomial functions on the space of jets can be easily extended to polynomial horizontal differential forms.

\vskip 3mm
\begin{proposition}\label{horizontal_polynomial} The following complexes are canonically isomorphic:
\beq\label{iso_poly_deRham_horizontal}
\left( \bar\Omega^{\bullet,l}_{\sst\mathrm{poly}}(\pi), \md_h \right)\simeq \left(\mathrm{Sym}^l\cD (\pi, \Lambda^{\bullet}_X), \md_{\sst\mathrm{DR}}\right)\,
\eeq
where $\bar\Omega^{\bullet,l}_{\sst\mathrm{poly}}(\pi)$ is a subcomplex of the horizontal complex $(\bar\Omega^{\bullet} (\pi), \md_h)$ consisting of horizontal differential forms which depend on jet variables as polynomials of the degree $l$.
\end{proposition}


\section{BRST cohomology in the space of local functionals.}\label{main}

\vskip 3mm\noindent Let $\eta$ be a $Q-$bundle over $T[1]X$, that is, a bundle in the category of $Q-$manifolds (cf.\cite{KS}), so that the $Q-$structure on the base is determined by the de Rham operator, regarded as a homological vector field. Apparently, not every section of $\eta$ in the graded sense is a section in the category of $Q-$manifolds, that is, not necessarily a $Q-$morphism; sections of $\eta$, which are $Q-$morphisms at the same time\footnote{Geometrically it means that those sections are tangent to the $Q-$stucture on the total space.}, are solutions to a certain system of PDEs. This system admits gauge symmetries (cf.\cite{HT}). The $Q-$stucture on the total space generates a homological vector field on the super space of sections $\underline{\Gamma}(\eta)$, denoted as $Q_{\sst\mathrm{BRST}}$; $\left(\underline{\Gamma}(\eta), Q_{\sst\mathrm{BRST}}\right)$ is the BV-BRST type model for the above system of PDEs.
$Q_{\sst\mathrm{BRST}}$ induces a nilpotent derivational of a (suitable) space of functionals $\mathscr{F}(\underline{\Gamma}(\eta))$; the problem is to compute the cohomology of the obtained complex.

\vskip 3mm\noindent In the case of a trivial bundle, the fiber of which is a PQ manifold, that is, a graded super symplectic manifold with a symplectic form of degree $\dim X-1$, so that the corresponding $Q-$field is Hamiltonian, we come to the classical BV theory for AKSZ type topological sigma models \cite{AKSZ}. In usual differential geometry, sections of a trivial bundle are in one-to-one correspondence with maps from the base to the fiber. Likewise, in the super case
\beq\label{iso_sections_functions} \underline{\Gamma}(\eta)\simeq \uHom (T[1]X, \cM)\,,\eeq
where $\cM$ is the fiber and $\uHom$ is the super space of maps. In general, the construction of $\uHom$ in (\ref{iso_sections_functions}) is rather complicated (cf.\cite{DM99} for the categorical approach; in \cite{BK}, $\uHom$ is explicitly represented by an infinite-dimensional supermanifold), unless the target is flat.

\vskip 3mm\noindent The choice of an appropriate space of functionals $\cF$ is not canonical. Furthermore, there is a tendency (even in non-super cases) to avoid possible troubles with an infinite-dimensional analysis by considering local (``jet depending") functionals in the sense of Section \ref{jet}. It seems to be at least equally useful for those theories which involve super maps. However, in TFTs the space of local functionals contains a very restrictive information, and we shall explicitly show that in the particular case of $\cM$ being a $\Z$-graded super vector space $L$ of finite type, i.e.
\beqn L^{\bullet}=\bigoplus\limits_{i\in\Z}L^{i}\eeq with $\mathrm{dim}L^{i}<\infty$ for all $i$, endowed with a structure of a Lie$_{\infty}$-algebra.

\vskip 3mm
\begin{deff} A Lie$_{\infty}$-algebra is a formal $Q$-manifold with the homological vector field vanishing at the origin.  \end{deff}

\begin{example} In the particular case of a Lie algebra the corresponding $Q$-manifold is $L:=\mathfrak{g}[1]$, where $\mathfrak{g}$ is the Lie agebra considered as a pure odd manifold, with $Q$-field given by the Chevalley-Eilenberg differential.
\end{example}

\begin{rem} In general, formal pointed (i.e. vanishing at the origin) $Q-$structures on $L$ are in one-to-one correspondence with nilpotent degree $1$ coderivations of
 the coalgebra $\mathrm{Sym}_c^+ (L)$, determined by an infinite sequence of maps $\mathrm{Sym}_c^i (L)\to L[1]$, $i\ge 1$. By use of the natural isomorphism
 $ \mathrm{Sym}^i (\g[1])\simeq \Lambda^i (\g)[i]$,
  we obtain a sequence of super skew-symmetric operations
  \beq\label{Lie_infinity}
  l_i\colon \Lambda^i (\g)\to \g [2-i]\, , \hspace{3mm} \forall i\ge 1\,,
  \eeq
   where $\g=L[-1]$; the latter was introduced under the name ``homotopy Lie algebras" \cite{SS79}.
\end{rem}

\noindent Denote with $\a^{\sst j,i}$ the bundle of differential $j-$forms on $X$ twisted by $L^i$, $\a^{\sst j,i}= \Lambda^j_X\otimes L^i$.

\vskip 3mm
\begin{lemma} \mbox{}
 \begin{enumerate}
  \item The super space of maps is given by $\Gamma (\a^{\bullet})$, where
   \beqn
    \a^{\bullet} =\bigoplus\limits_{q\in\Z} \a^{\sst q}\, , \hspace{3mm} \a^{\sst q} =\bigoplus\limits_{j=0}^{\dim X} \a^{\sst j,q-j}
   \eeq
  is regarded as a $\Z-$graded super vector bundle with the total $\Z-$grading induced by the degree of forms and the grading in $L$.
  \item $Q_{\sst\mathrm{BRST}}=Q_{\sst\mathrm{DR}}+ Q_L$, where $Q_{\sst\mathrm{DR}}$ is a (linear) homological vector field,
  given by the $L-$twisted de Rham operator, while $Q_L$ is a pointed formal $Q-$field, determined by the super multi-linear over $\Omega^{\bullet} (X)$ extension of the coderivation of $\mathrm{Sym}_c^+ (L)$.
 \end{enumerate}
\end{lemma}

\noindent As it was previously mentioned, the choice of $\mathscr{F}$, the space of functionals, is not canonical. On the other hand, the super space of maps is now represented by sections of graded super vector bundle over an even (``bosonic") base $X$. One may address the naturally looking question of computing the cohomology in the space of local functionals, which are polynomials in jet variables, with respect to the differential $\mathrm{\bf s}$, where $\mathrm{\bf s}$ is the evolutionary vector field corresponding to $Q_{\sst\mathrm{BRST}}$. In other words, we are interested in $H^{\bullet,n}(\mathrm{\bf s}\mid \md_h)$, where
\beqn
 H^{\bullet,n}(\mathrm{\bf s}\mid \md_h)\colon =\bigoplus\limits_{g\in\Z} H^g \left( H^n \left( \bar\Omega^{\bullet,\bullet}_{\sst\mathrm{poly}}(\pi), \md_h\right), \mathrm{\bf s} \right),
\eeq
($\omega^{n,g}\in \Omega^{n,g}_{\sst\mathrm{poly}}(\pi)$ is a $n$-horizontal form of $g\in \Z$ degree).

\vskip 3mm
\begin{proposition}\label{2d_term_iso_total} One has $H^{g,n}(\mathrm{\bf s}\mid \md_h)\simeq H^{g+n} (\mathrm{\bf s}+\md_h)$.
\end{proposition}

\vskip 3mm\noindent\proof We apply the canonical isomorphism (\ref{iso_poly_deRham_horizontal}). Let us consider the corresponding bicomplex

\beq\label{bicomplex1}
\xymatrix{
 &\vdots &  & \vdots\\
0 \ar[r] &\left[\mathrm{Sym}^+\mathcal{D}(\a^{\bullet},\mathbbold{1})\right]^{g+1} \ar[r]^-{\md_{\sst\mathrm{DR}}}\ar[u]^{\mathrm{\bf{s}}}&...\ar[r]^-{\md_{\sst\mathrm{DR}}}&\ar[u]^{\mathrm{\bf{s}}}\left[\mathrm{Sym}^+\mathcal{D}(\a^{\bullet},\Lambda^n_X)\right]^{g+1}\ar[r]&0\\
0 \ar[r]&\left[\mathrm{Sym}^+\mathcal{D}(\a^{\bullet},\mathbbold{1})\right]^{g}\ar[r]^-{\md_{\sst\mathrm{DR}}}\ar[u]^{\mathrm{\bf{s}}}&
...\ar[r]^-{\md_{\sst\mathrm{DR}}}&\ar[u]^{\mathrm{\bf{s}}}\left[\mathrm{Sym}^+\mathcal{D}(\a^{\bullet},\Lambda^n_X)\right]^{g}\ar[r]&0.\\
& \vdots\ar[u]^{\mathrm{\bf{s}}}& &\vdots \ar[u]^{\mathrm{\bf{s}}} }
\eeq

\noindent We examine the spectral sequence determined by (\ref{bicomplex1}), where the filtration is chosen such that the cohomology of the rows are to be taken at first. The $E_1-$term of the above spectral sequence can be computed by use of the following Lemma.

\vskip 3mm\begin{lemma}\label{horizontal_cohomology_for_polynomial}
Let $\a$ be a vector bundle. Then one has for all $l>0$
\beq
 H^i \left(\mathrm{Sym}^l\cD (\a, \Lambda^{\bullet}_X), \md_{\sst\mathrm{DR}}\right)=\left\{
 \be{cc}
  \mathrm{Sym}^{l-1}_{\sst\mathrm{self}}\cD (\a, \hat\a) , & i=n \\
  0, & i<n
 \ee
 \right.
\eeq
where $\mathrm{Sym}^{l-1}_{\sst\mathrm{self}}\cD (\a, \hat\a)$ is the space the $(q-1)-$linear symmetric differential operators, (formally) self-adjoint with respect to each argument. In particular, for $q=1$ one has
\beq
 H^i \left(\cD (\a, \Lambda^{\bullet}_X), \md_{\sst\mathrm{DR}}\right)=\left\{
 \be{cc}
  \hat\a , & i=n \\
  0, & i<n
 \ee
 \right.
 \eeq
\end{lemma}

\noindent The proof is rather standard; we notice that the differential in the above complex commutes with the right $\cO_X-$action coming from the $\cO_X-$module structure on $\Gamma(\a)$, thus one has a complex of locally trivial $\cO_X-$modules or, equivalently, a complex of vector bundle morphisms. This implies that formula (\ref{horizontal_cohomology_for_polynomial}) can be derived in any local coordinates, using the symbolic filtration. A similar result, involving $\cC-$differential operators instead of $\cD_X$, is obtained in the case of the Vinogradov's $\cC-$spectral sequence (cf.\cite{Boc97,KV98}). Taking into account that the $E_1-$term is concentrated in degree $n$ only, we immediately obtain that the above spectral sequence converges in the second term, thus the second term of the spectral sequence is isomorphic to the cohomology of the total complex with the differential $\mathrm{\bf s}+\md_h$.
 Given that the second term of the spectral sequence is nothing but $H^{g,n}(\mathrm{\bf s}\mid \md_h)$, we complete the proof of Proposition \ref{2d_term_iso_total}.
$\square$

\begin{theorem}\label{main_th} One has $H^{g,n}(\mathrm{\bf{s}}\mid \md_h)\simeq \left( H^{\bullet}_{\mathrm{DR}} (X)\otimes H^{\bullet}_Q (L)\right)^{g+n}$.
\end{theorem}

\vskip 3mm\noindent\proof
The differential, given by the evolutionary vector field $\mathrm{\bf s}$, splits into the two parts $\mathrm{\bf s} = \cQ_L+\de_{\sst\mathrm{DR}}$, which come from $Q_L$ and $Q_{\sst\mathrm{DR}}$, respectively. In particular $\de_{\sst\mathrm{DR}}$ is the derivation of $\mathrm{Sym}^+\mathcal{D}(\a^{\bullet},\Lambda^p_X)$ induced by the right composition of $Q_{\sst\mathrm{DR}}$ with differential operators $\mathcal{D}(\a^{\bullet},\Lambda^p_X)$. The two independent gradings allow to define another bicomplex

\beq\label{bicomplex2}
\xymatrix{
&0 &  0\\
 \ldots\ar[r]^-{\cQ_L} &\left[\mathrm{Sym}^+\mathcal{D}(\a^{\bullet},\Lambda^p_X)\right]^{0,i}
\ar[u]\ar[r]^-{\cQ_L}\ar[r]^-{\cQ_L}&\ar[u]
\left[\mathrm{Sym}^+\mathcal{D}(\a^{\bullet},\Lambda^p_X)\right]^{0,i+1}\ar[r]^-{\cQ_L}&\ldots\\
 & \vdots\ar[u]^{\mathrm{\de_{\sst\mathrm{DR}}}} & \vdots \ar[u]^{\mathrm{\de_{\sst\mathrm{DR}}}}\\
 \ldots\ar[r]^-{\cQ_L} &\left[\mathrm{Sym}^+\mathcal{D}(\a^{\bullet},\Lambda^p_X)\right]^{-j+1,i}
\ar[u]^{\mathrm{\de_{\sst\mathrm{DR}}}}\ar[r]^-{\cQ_L}\ar[r]^-{\cQ_L}&\ar[u]^{\mathrm{\de_{\sst\mathrm{DR}}}}
\left[\mathrm{Sym}^+\mathcal{D}(\a^{\bullet},\Lambda^p_X)\right]^{-j+1,i+1}\ar[r]^-{\cQ_L}&\ldots\\
\ldots\ar[r]^-{\cQ_L}&\left[\mathrm{Sym}^+\mathcal{D}(\a^{\bullet},\Lambda^p_X)\right]^{-j,i}
\ar[u]^{\mathrm{\de_{\sst\mathrm{DR}}}}\ar[r]^-{\cQ_L}\ar[r]^-{\cQ_L}&\ar[u]^{\mathrm{\de_{\sst\mathrm{DR}}}}
\left[\mathrm{Sym}^+\mathcal{D}(\a^{\bullet},\Lambda^p_X)\right]^{-j,i+1}\ar[r]^-{\cQ_L}&\ldots.\\
& \vdots\ar[u]^{\mathrm{\de_{\sst\mathrm{DR}}}} & \vdots \ar[u]^{\mathrm{\de_{\sst\mathrm{DR}}}} }
\eeq

\noindent Here $\mathrm{Sym}^+\mathcal{D}(\a^{\bullet},\Lambda^p_X)$ is a canonically bi-graded vector space, such that, in particular, the first degree, corresponding to the one in $\a$, is always non-positive. Furthermore, we are finally interested in the calculation of the total cohomology $\md_{\sst\mathrm{DR}}+\mathrm{\bf s}$ (due to Proposition \ref{2d_term_iso_total}) which is made up by three differentials $\md_{\sst\mathrm{DR}}$, $\cQ_{L}$, and $\de_{\sst\mathrm{DR}}$ with three independent gradings.
We combine the fist two of them and construct a filtration for the bicomplex $(\mathrm{Sym}^+\mathcal{D}(\a^{\bullet},\Lambda^\bullet_X), \cQ_{L}+ \md_{\sst\mathrm{DR}}, \de_{\sst\mathrm{DR}})$, such that the cohomology with respect to $\de_{\sst\mathrm{DR}}$  are to be computed at first.
Thus we need to calculate the cohomology of the columns in (\ref{bicomplex2}). We notice that the following complex is formally exact (in the sense of \cite{Quill,Spe69})
$$
\xymatrix{
\a^{0,i}\ar[r]^-{Q_{\sst\mathrm{DR}}}&
\a^{1,i} \ar[r]^-{Q_{\sst\mathrm{DR}}}
& \ldots \ar[r]^-{Q_{\sst\mathrm{DR}}}&\a^{n,i}\ar[r] &0.\\}
$$
In particular this means that applying to this complex the jet infinity functor, $\mathrm{J}^{\infty}$, we get an exact sequence of $\cO_{X}$-modules
$$
\xymatrix{
\mathrm{J}^{\infty}(\a^{0,i})\ar[r]&
\mathrm{J}^{\infty}(\a^{1,i}) \ar[r]
& \ldots \ar[r]&\mathrm{J}^{\infty}(\a^{n,i})\ar[r] &0;\\}
$$
we extend it to the following exact sequence of $\cO_{X}$-modules
$$
\xymatrix{
\Lambda^{n-p}_X\otimes\mathrm{J}^{\infty}(\a^{0,i})\ar[r]&
\Lambda^{n-p}_X\otimes\mathrm{J}^{\infty}(\a^{1,i}) \ar[r]
& \ldots \ar[r]&\Lambda^{n-p}_X\otimes\mathrm{J}^{\infty}(\a^{n,i})\ar[r] &0,\\}
$$
(for $0\leq p\leq n$) where the horizontal arrows are still induced by the operator ${Q_{\sst\mathrm{DR}}}$.
Dualizing the previous sequence, by the use of the left-exact contravariant $\Hom(-,\Lambda^{n}_X)$ functor, we get a sequence of right $\cO_X$-modules which is exact everywhere except at the zero spot (i.e. it is a resolution of a cokernel)

\beq\label{de_Rham_resolution_linear}
\xymatrix{
0\ar[r]&\mathcal{D}(\a^{n,i},\Lambda^p_X)
\ar[r]^-{\mathrm{\de_{\sst\mathrm{DR}}}}&
\mathcal{D}(\a^{n-1,i},\Lambda^p_X)
\ar[r]^-{\mathrm{\de_{\sst\mathrm{DR}}}}&
\ldots\ar[r]^-{\mathrm{\de_{\sst\mathrm{DR}}}}&\mathcal{D}(\a^{0,i},\Lambda^p_X)
\ar[r]&\mathrm{coker}\ar[r]& 0 .}
\eeq

\vskip 3mm\noindent More precisely, it means that

\beqn
 H^j \left(\mathcal{D}(\a^{\bullet,i},\Lambda^p_X), \de_{\sst\mathrm{DR}}\right)=\left\{
 \be{cc}
  \Hom (L^i, \Lambda^p_X) , & j=0 \\
  0, & j<0.
 \ee
 \right.
\eeq

\vskip 3mm\noindent Now we take the symmetric powers of (\ref{de_Rham_resolution_linear}) and we get

\beq\label{cohomology}
 H^j \left(\left[\mathrm{Sym}^+\mathcal{D}(\a^{\bullet},\Lambda^p_X)\right]^{\bullet,i}, \de_{\sst\mathrm{DR}}\right)=\left\{
 \be{cc}
  \left[\mathrm{Sym}^+ (L^*)\right]^i\otimes \Omega^p (X)  , & j=0 \\
  0, & j<0.
 \ee
 \right.
\eeq

\noindent This completes the calculation of the term $E_1$.
 The second term of the above spectral sequence coincides with the total cohomology
of the bicomplex $(\Omega^\bullet (X)\otimes \left[\mathrm{Sym}^+ (L^*)\right]^\bullet, \md_{\sst\mathrm{DR}}, \cQ_{L})$, which is simply the tensor product of
$(\Omega^\bullet (X), \md_{\sst\mathrm{DR}})$ and
$(\left[\mathrm{Sym}^+ (L^*)\right]^\bullet, \cQ_{L})$. Thus we have the following K\"{u}nneth type formula (cf.\cite{Rot})
\beq\label{last_second_term}
H^{p} \Big(\Omega^\bullet (X)\otimes \left[\mathrm{Sym}^+ (L^*)\right]^\bullet , \md_{\sst\mathrm{DR}}+\cQ_{L}\Big)= \bigoplus\limits_{i+j=p}
H^i_{\mathrm{DR}} (X)\otimes H^j_Q (L)\,.
\eeq

\noindent
 We observe that the bicomplex associated to the couple $(\md_{\sst\mathrm{DR}}+\cQ_{L},\de_{\sst\mathrm{DR}})$ verifies the hypothesis of Remark $\ref{spectral_seq}$ below, in view of Lemma $\ref{horizontal_cohomology_for_polynomial}$ and eq. (\ref{cohomology}).
 Therefore, the $E_2-$term of the associated spectral sequence coincides with the total cohomology with the differential $\md_{\sst\mathrm{DR}}+\cQ_{L}+\de_{\sst\mathrm{DR}}$ and thus,
  using (\ref{last_second_term}), we accomplish the proof of Theorem \ref{main_th}. $\square$

\vskip 3mm\noindent The previous proof contains a result which can be stated in all generality in the following way.
\vskip 3mm\begin{lemma}\label{de_Rham_resolution}
Let $\b$ be a vector bundle and $\gamma$ be another vector space endowed with a flat connection.
 Consider the following complex of left $\cO_X-$modules $\left(\cD (\Lambda_X^{\bullet}\otimes \gamma, \b), \de_{\sst\mathrm{DR}}\right)$,
where the differential $\de_{\sst\mathrm{DR}}$ is induced by the right composition with the de Rham operator twisted by the flat connection in $\gamma$.
Then one has
\beq
 H^i \left(\mathrm{Sym}^+\cD (\Lambda_X^{\bullet}\otimes \gamma, \b), \de_{\sst\mathrm{DR}}\right)=\left\{
 \be{cc}
 \Gamma \left(\mathrm{Sym}^+ (\gamma^*)\otimes \b \right) , & i=0 \\
  0, & i<0
 \ee
 \right.
\eeq
\end{lemma}
\vskip 3mm
\begin{rem}\label{spectral_seq}
Let $K^{\bullet}$ be the total complex of a bicomplex $K^{\bullet,\bullet}$ with linear maps
\beqn d^{1}\colon K^{p,q}\to K^{p+1,q}\, , \hspace{3mm} d_{2}\colon K^{p,q}\to K^{p,q+1}\,,\eeq such that $(d_{1})^{2}=0,\;(d_{2})^{2}=0$ and
$d_{2}d_{1}+d_{1}d_{2}=0$.
There are two filtrations
\beqn K^i_p(1)=\bigoplus\limits_{j+q=i,\; j\geq p}K^{j,p}\, , \hspace{3mm} K^i_q(2)=\bigoplus\limits_{p+j=i, \; j\geq q}K^{p,j}\,.\eeq
These two filtrations yield two spectral sequences, denoted respectively by $E_{r}^{p,q}(1),\;E_{r}^{p,q}(2)$; in particular recall that
$E_{2}^{p,q}(1)=H_{1}^p \left( H_{2}^q(K^{\bullet,\bullet})\right)$ and $E_{2}^{p,q}(2)=H_{2}^p \left( H_{1}^q(K^{\bullet,\bullet})\right)$.
Now assume that both filtrations are regular. In this case both spectral sequence converge to the common limit $H^{\bullet}(K^{\bullet})$.
\vskip 3mm\noindent Suppose that in the following diagram

$$\xymatrix{0&\ar[l]K^{2,0}\ar[d]^{d_{1}}&\ar[l]^-{d_{2}}\ar[d]^{d_{1}}K^{2,1}&\ar[l]^-{d_{2}}\ar[d]^{d_{1}}K^{2,2}&\ar[l]\\
0&\ar[l]K^{1,0}\ar[d]^{d_{1}}&\ar[l]^-{d_{2}}\ar[d]^{d_{1}}K^{1,1}&\ar[l]^-{d_{2}}\ar[d]^{d_{1}}K^{2,2}&\ar[l]\\
0&\ar[l]K^{0,0}\ar[d]^{}&\ar[l]^-{d_{2}}\ar[d]^{}K^{0,1}&\ar[l]^-{d_{2}}\ar[d]^{}K^{0,2}&\ar[l]\\
&0&0&0&\\} $$
all the sequences are exact except for the terms in the left column and
bottom row. We have two complexes $Q_{1}^{\bullet}$ and $Q_{2}^{\bullet}$, where $Q_{1}^{i}=H^{0}(K^{i,\bullet},d_{2})$ and $Q_{2}^{i}=H^{0}(K^{\bullet,i},d_{1})$ and the differentials are induced by $d_{1}$ and $d_{2}$ respectively. It follows that $E_{2}^{p,q}(1)=E_{3}^{p,q}(1)=\ldots=E_{\infty}^{p,q}(1)$ is equal to $H^{p}(Q_{1}^{\bullet})$ (if $q=0$ and zero otherwise) and $E_{2}^{p,q}(2)=E_{3}^{p,q}(2)=\ldots=E_{\infty}^{p,q}(2)$ is equal to $H^{q}(Q_{2}^{\bullet})$ (if $p=0$ and zero otherwise). Since both spectral sequences converge to a common limit, we conclude that $H^{i}(Q_{1}^{\bullet})=H^{i}(Q_{2}^{\bullet})$.
\end{rem}

\vskip 5mm\noindent Giuseppe Bonavolont\`a\\
 Mathematics Research Unit\\
University of Luxembourg \\
6, rue R. Coudenhove-Kalergi\\
L-1359 Luxembourg City
\vskip 5mm\noindent Alexei Kotov\\
Department of Mathematics and Statistics\\
Faculty of Science and Technology\\
University of Troms{\o},
N-9037 Troms{\o}

\end{document}